\documentclass[]{article}
\usepackage{proceed2e}

\usepackage{times}

\usepackage{times}

\usepackage{graphicx} 

\usepackage{subfigure} 

\usepackage[round]{natbib}

\usepackage{algorithm}
\usepackage{bm}
\usepackage{amssymb,amsmath}

\usepackage{hyperref}
\usepackage{color}

\usepackage{algpseudocode}

\usepackage{amsfonts}
\newtheorem{theorem}{Theorem}

\newtheorem{definition}{Definition}
\newcommand{\BEQ}{\begin{equation}}
\newcommand{\EEQ}{\end{equation}}
\newcommand{\BEA}{\begin{eqnarray}}
\newcommand{\EEA}{\end{eqnarray}}
\newcommand{\BGA}{\begin{gather}}
\newcommand{\EGA}{\end{gather}}

\newcommand{\vx}{{\mathbf x}}

\newcommand{\vy}{{\mathbf y}}
\newcommand{\vu}{{\mathbf u}}

\newcommand{\vh}{{\mathbf H}}
\newcommand{\ii}{{(i)}}

\newcommand{\comment}[1]{}

\DeclareMathOperator*{\argmax}{arg\,max}

\title{Estimating Mutual Information by Local Gaussian Approximation}

\author{ {\bf Shuyang Gao} \\
Information Sciences Institute \\
University of Southern California\\
{sgao@isi.edu}
\And
{\bf Greg {Ver Steeg}}  \\
Information Sciences Institute \\
University of Southern California\\
{gregv@isi.edu}
\And
{\bf Aram Galstyan}   \\
Information Sciences Institute \\
University of Southern California\\
{galstyan@isi.edu}
}

\begin{document} 
\maketitle

\begin{abstract} 
Estimating mutual information (MI) from samples is a fundamental problem  in statistics, machine learning, and data analysis. Recently it was shown that a popular class of non-parametric MI estimators perform very poorly for strongly dependent variables and have sample complexity that scales exponentially with the true MI. This undesired behavior was attributed to the reliance of those estimators on local uniformity of the underlying (and unknown) probability density function. Here we present a novel semi-parametric estimator of mutual information, where at each sample point, densities are {\em locally} approximated by a Gaussians distribution. We demonstrate that the estimator is asymptotically unbiased. We also show that  the proposed estimator has a superior performance compared to several baselines, and is able to accurately measure relationship strengths over many orders of magnitude. 

\end{abstract}

\section{Introduction} \label{sec:intro}

Mutual information (MI) is a fundamental measure of dependence between two random variables. While it initially arose in the theory of communication as a natural measure of ability to communicate over noisy channels~\citep{shannon}, mutual information has since been used in different disciplines such as machine learning, information retrieval, neuroscience, and computational biology, to name a few. This widespread use is due in part to the generality of the measure, which allows it to characterize dependency strength for both linear and non-linear relationships between arbitrary random variables. 

Let us consider the following basic problem, where, given a set of i.i.d.\ samples from an unknown, absolutely continuous joint distribution, our goal is to estimate the mutual information from these samples. A naive method would be first to learn the underlying probability distribution using either parametric or non-parametric methods, and then calculate the mutual information from the obtained distribution. Unfortunately, this naive approach often fails, as it requires a very large number of samples, especially in high dimensions. A different approach is to estimate mutual information directly from samples. For instance, rather than estimating the whole probability distribution, one could estimate the density (and its marginals) only at each sample point, and then plug those estimates into the expression for mutual information. This type of direct estimators been shown to be a more feasible method for estimating MI in higher dimensions. An important and very popular class of such estimators is based on  k-nearest-neighbor (kNN) graphs and their generalizations~\citep{kNN_naive,kraskov,GNNG}.

Despite the widespread popularity of the direct estimators, it was recently demonstrated that those methods fail to accurately estimate mutual information for \textit{strongly dependent} variables~\citep{shuyang2015AISTATS}. Specifically, it was shown that accurate estimation of mutual information between two strongly dependent variables requires a number of samples that scales {\em exponentially} with the true mutual information. This undesired behavior was contributed to the assumption of local uniformity of the underlying distribution postulated by those estimators. To address this shortcoming,  ~\citep{shuyang2015AISTATS}  proposed to add a correction term to compensate for non-uniformity, based on local PCA-induced neighborhoods. Although intuitive, the resulting estimator relied on a heuristically tuned threshold parameter  and  had no theoretical performance guarantees~\citep{shuyang2015AISTATS}.

Our main contribution is to propose a novel mutual information estimator based on \textit{local Gaussian approximation}, with provable performance guarantees, and superior empirical performance compared to existing estimators over a wide range of relationship strength.
Instead of assuming a uniform distribution in the local neighborhood, our new estimator assumes a Gaussian distribution \textit{locally} around each point. The new estimator leverages previous results on \textit{local likelihood density estimation}~\citep{hjort1996locally,loader1996local}. As our main theoretical result, we demonstrate that the new estimator is asymptotically unbiased. We also demonstrate that the proposed estimator performs as well as existing  baseline estimators for weak relationships, but outperforms all of those estimators for stronger relationships. 

The paper is organized as follows. In the next section, we review the basic definitions of information-theoretic concepts such as mutual information and  formally define our problem. In section~\ref{sec:limit}, we review the limitations of current mutual information estimators as pointed out in \citep{shuyang2015AISTATS}. Section~\ref{sec:lld} introduces local likelihood density estimation. In Section~\ref{sec:lge} we use this density estimator to propose a novel entropy and mutual information estimator, and summarize certain theoretical properties of those estimator, which are then proved in Section~\ref{sec:proof}. Section~\ref{sec:exp} provides numerical experiments demonstrating the superiority of the proposed estimator. We conclude the paper with a brief survey of related work followed by the discussion of our main results and some open problems.

\section{Formal Problem Definition}

In this section we briefly review the formal definition of Shannon entropy and mutual information, before formally defining the objective of our paper.
\begin{definition}\label{def:entropy_def}
Let $\vx$ denote a \textsl{d}-dimensional absolutely continuous random variable with probability density function $f: \mathbb R^d \to \mathbb R$. The \textsl{Shannon differential entropy} is defined as
\BEQ
H\left( {\mathbf{x}} \right) =  - \int\limits_{{\mathbb R^d}} {f\left( {\mathbf{x}} \right)\log f\left( {\mathbf{x}} \right)d{\mathbf{x}}} 
\label{eq:entropy}
\EEQ
\end{definition}

\begin{definition}\label{def:mi_def}
Let $\vx$ and $\vy$ denote  \textsl{d}-dimensional and  \textsl{b}-dimensional absolutely continuous random variables with probability density function $f_X: \mathbb R^d \to \mathbb R$ and $f_Y: \mathbb R^b \to \mathbb R$, respectively. Let $f_{XY}$ denote the joint probability density function of $\vx$ and $\vy$. The \textsl{mutual information} between $\vx$ and $\vy$ is defined as
\BEQ
I\left( {{\mathbf{x}}:{\mathbf{y}}} \right) = \int\limits_{{\mathbf{y}} \in {\mathbb R^b}} {\int\limits_{{\mathbf{x}} \in {\mathbb R^d}} {{f_{XY}}\left( {{\mathbf{x}},{\mathbf{y}}} \right)\log \frac{{{f_{XY}}\left( {{\mathbf{x}},{\mathbf{y}}} \right)}}{{{f_X}\left( {\mathbf{x}} \right){f_Y}\left( {\mathbf{y}} \right)}}d{\mathbf{x}}d{\mathbf{y}}} } 
\EEQ
\end{definition}
It is easy to show that
\BEQ \label{eq:ent_mi}
I\left( {{\mathbf{x}}:{\mathbf{y}}} \right) = H\left( {\mathbf{x}} \right) + H\left( {\mathbf{y}} \right) - H\left( {{\mathbf{x}},{\mathbf{y}}} \right) \ , \
\EEQ
where $H(\vx,\vy)$ stands for the joint entropy of $(\vx, \vy)$, and can be calculated from Eq.~\ref{eq:entropy} using the joint density $f_{XY}$. We use the natural logarithms so that information is measured in nats. 

It is sometime useful to represent entropy and mutual information as the following expectations:
\BEA
H\left( {\mathbf{x}} \right) &=&  \mathbb{E}_{X} [- \log f({\vx})]\label{eq:naiveE} \\
I\left( {{\mathbf{x}}:{\mathbf{y}}} \right) &=&  \mathbb{E}_{XY} \biggl [ \log \frac{{{f_{XY}}\left( {{\mathbf{x}},{\mathbf{y}}} \right)}}{{{f_X}\left( {\mathbf{x}} \right){f_Y}\left( {\mathbf{y}} \right)}} \biggr ] \label{eq:naiveI} 
\EEA
Assume now we are given $N$ i.i.d.\ samples $(\mathcal{X},\mathcal{Y})=\{{(\vx,\vy)^\ii}\}^n_{i=1}$ from the unknown joint distribution $f_{XY}$. Our goal is then to construct a mutual information estimator $\hat{I}(\vx:\vy)$ based on those samples.

\section{Limitations of Nonparametric MI Estimators} \label{sec:limit}

\begin{figure}[htbp] 
   \centering
   \subfigure[]{\includegraphics[width=0.2\textwidth]{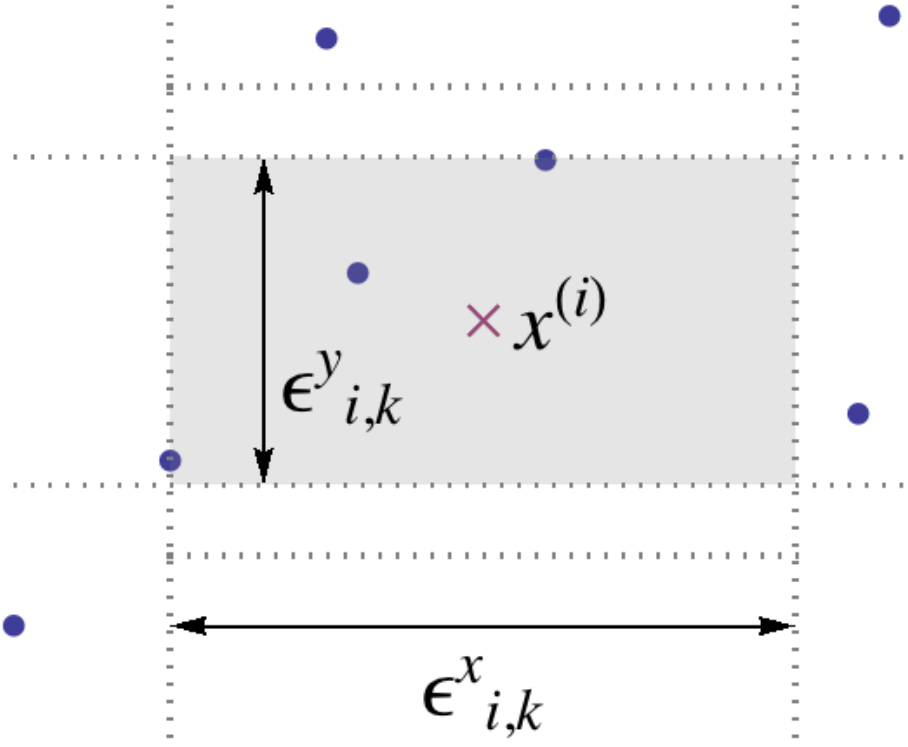} \label{fig:ksg}} 
   \quad
   \subfigure[]{ \raisebox{0.22\height}{\includegraphics[width=0.2\textwidth]{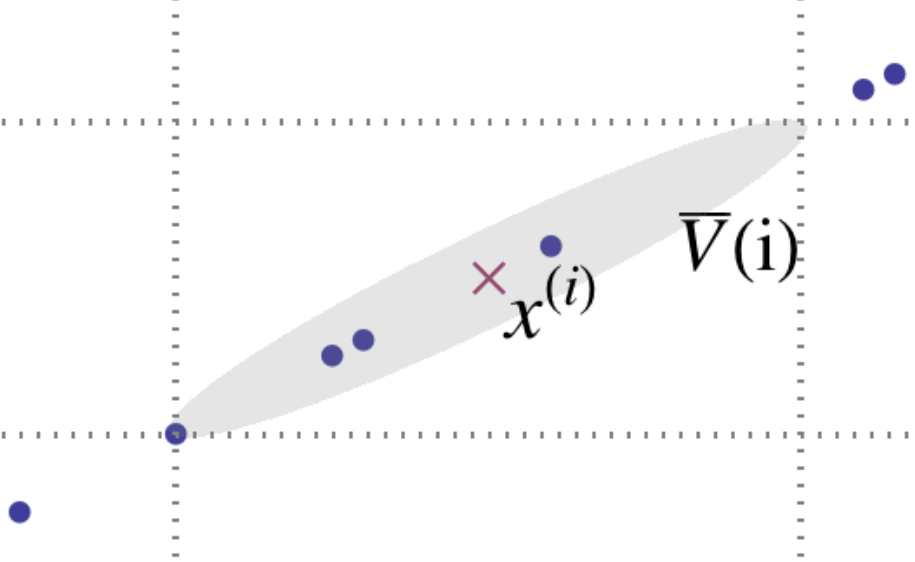} \label{fig:ksg_problem}} } 
   \caption{For a given sample point $\vx^\ii$, we show the max-norm rectangle containing $k$ nearest neighbors (a) for points drawn from a uniform distribution, $k=3$, (shaded area), and (b) for points drawn from a distribution over two strongly correlated variables, $k=4$, (the area within dotted lines). }
\label{fig:1}
\end{figure}

As pointed out in Section~\ref{sec:intro}, one of the most popular class of mutual information estimators is based on k-nearest neighbor (kNN) graphs and their generalizations~\citep{kNN_naive,kraskov,GNNG}. However, it was recently shown that for strongly dependent variables, those estimators tend to underestimate the mutual information~\citep{shuyang2015AISTATS}. To understand this problem, let us focus on kNN-based estimator as an example. The kNN estimator assumes uniform density within the kNN rectangle (containing k-nearest neighbors), as shown in Figure~\ref{fig:ksg}. Generally speaking, this assumption can be made valid for any relationship as long as we have sufficient number of samples. However, for limited sample size, this assumption becomes problematic when the relationship between the two variables becomes sufficiently strong. In fact, as shown in~Fig.~\ref{fig:ksg_problem}, the obtained \textit{local} neighborhood induced by kNN is beyond the support of the probability distribution (shaded area).

This undesired behavior is closely related to the so-called {\em boundary effect} that occurs in nonparametric density estimation problem. Namely, for strongly dependent random variables, almost all the sample points are close to the boundary of the support (as illustrated in Figure~\ref{fig:ksg_problem}), making the density estimation problem difficult. 

To relax the local uniformity assumption in kNN-based estimators, \citep{shuyang2015AISTATS} proposed to replace the axis-aligned rectangle with a PCA-aligned rectangle \textit{locally}, and use the volume of this rectangle for estimating the unknown density at a given point. Mathematically, the above revision was implemented by introducing a novel term that accounted for local non-uniformity. It was shown the the revised estimator significantly outperformed the existing  estimators for strongly dependent variables. Nevertheless, the estimator suggested in ~\citep{shuyang2015AISTATS} relied on a heuristic for determining when to use the correction term, and did not have any theoretical guarantees. In the remaining of this paper, we suggest a novel estimator based on \textit{local gaussian approximation}, as more general approach to overcome the above limitations. The main idea is that, instead of assuming a uniform distribution around the local kNN-  or a PCA-aligned rectangle, we approximate the unknown density at each sample point by a local \textit{Gaussian} distribution, which is estimated using the k-nearest neighborhood of that point. In addition to demonstrating superior empirical performance of the proposed estimator, we also show that it is asymptotically unbiased.

\section{Local Gaussian Density Estimation} \label{sec:lld}

In this section, we introduce a density estimation method called local Gaussian density estimation, or LGDE~\citep{hjort1996locally}, which serves as the basic building block for the proposed mutual information estimator.

Consider $N$ i.i.d.\ samples $\vx_1, \vx_2, ..., \vx_N$ drawn from an unknown density $f(\vx)$, where $\vx$ is a $d$-dimensional continuous random variable. The central idea behind LGDE is to locally approximate the unknown probability density at point $\vx$ using a Gaussian parametric family ${\mathcal{N}_d}\left( {\bm{\mu(\vx)} ,\bm{\Sigma(\vx)} } \right)$, where $\bm{\mu(\vx)}$  and $\bm{\Sigma}(\vx)$ are the ($\vx$-dependent) mean and covariance matrix of each local approximation. This intuition is formalized in the following definition:
\begin{definition}[Local Gaussian Density Estimator]\label{def:lld} Let $\vx$ denote a $d$-dimensional absolutely continuous random variable with probability density function  $f(\vx)$, and let  $\{\vx_1$, $\vx_2$,..., $\vx_N\}$ be $N$ i.i.d. samples drawn from $f(\vx)$. Furthermore, let  $K_\vh(\vx)$ be a product kernel with diagonal bandwidth matrix $\vh = diag(h_1,h_2,...,h_d)$, so that ${{\mathbf{K}}_{\mathbf{H}}}\left( {\mathbf{x}} \right) = h_1^{ - 1}K\left( {h_1^{ - 1}{x_1}} \right)h_2^{ - 1}K\left( {h_2^{ - 1}{x_2}} \right)...h_d^{ - 1}K\left( {h_d^{ - 1}{x_d}} \right)$, where $K(\cdot)$ can be any one-dimensional kernel function. Then the  Local Gaussian Density Estimator, or LGDE, of $f(\vx)$ is given by 
\BEQ
\widehat f\left( {\mathbf{x}} \right) = \mathcal{N}_d\left( {{\mathbf{x}};{\bm{\mu}(\vx) ,\bm{\Sigma}(\vx) }} \right) \ , 
\EEQ
Here $\bm{\mu}, \bm{\Sigma}$ are different for each point $\vx$, and are obtained by solving the following optimization problem,
\BEQ
\bm{\mu}(\vx),\bm{\Sigma}(\vx)  = \argmax _{\bm{\mu}, \bm{\Sigma} }\mathcal L\left( {{\mathbf{x}},\mu,\Sigma} \right) \ , 
\EEQ
where $\mathcal{L}\left( {{\mathbf{x}},\bm{\mu} ,\bm{\Sigma} } \right)$  is the \textit{local likelihood function} defined as follows:
\BEA \label{eq:llf}
\mathcal{L}\left( {{\mathbf{x}},\bm{\mu} ,\bm{\Sigma} } \right) &=& \frac{1}{N}\sum\limits_{i = 1}^N {{\mathbf{K}_\vh}\left( {{{\mathbf{x}}_i} - {\mathbf{x}}} \right)\log } \mathcal{N}_d\left( {{{\mathbf{x}}_i};\bm{\mu} ,\bm{\Sigma} } \right) \nonumber \\
&-&\int {{\mathbf{K}_\vh}\left( {{\mathbf{t}} - {\mathbf{x}}} \right)\mathcal{N}_d\left( {{\mathbf{t}};\bm{\mu} ,\bm{\Sigma} } \right)d{\mathbf{t}}} 
\EEA
\label{def:LGDE}
\end{definition}
The first term in the right hand side of Eq.~\ref{eq:llf} is the localized version of Gaussian log-likelihood. One can see that without the kernel function, Eq.~\ref{eq:llf} becomes similar to the global log-likelihood function of the Gaussian parametric family. However, since we do not have sufficient information to specify a global distribution, we make a local smoothness assumption by adding this kernel function. The second term of right hand side in Eq.~\ref{eq:llf} is a penalty term to ensure the consistency of the density estimator.

The key difference between kNN density estimator and LGDE is that the former assumes that the density is  locally uniform over the neighborhood of each sample point, whereas the latter method relaxes \textit{local uniformity} to \textit{local linearity}\footnote{To elaborate on the local linearity, we note that Gaussian distribution is essentially a special case of Elliptical distribution $f(\vx)=k*g((\vx-\bm\mu)^T{\bm\Sigma}^{-1}(\vx-\bm\mu))$. Therefore, the local Gaussian approximation actually assumes a rotated hyper-ellipsoid locally at each point.}, which allows to compensates for the boundary bias. In fact, any non-uniform parametric probability distribution is suitable for fitting a local distribution under the local likelihood, and the Gaussian distribution used here is simply one realization. 

Theorem~\ref{theo:lgd_consist} below establishes the consistency property of this local Gaussian estimator; for a detailed proof see~\citep{hjort1996locally}. 

\begin{theorem}[~\citep{hjort1996locally} ]\label{theo:lgd_consist}
Let $\vx$ denote a $d$-dimensional absolutely continuous random variable with probability density function  $f(\vx)$, and let  $\{\vx_1$, $\vx_2$,..., $\vx_N\}$ be $N$ i.i.d. samples drawn from $f(\vx)$. Let $\widehat f\left( x \right)$  be the Local Gaussian Density Estimator with diagonal bandwidth matrix $diag(h_1, h_2,...,h_d)$, where the diagonal elements $h_i$-s satisfy the following conditions:
\BEQ
\lim_{N\to \infty} h_i = 0 \ ,  \lim_{N\to \infty} N h_i = \infty , i=1,2,\dots, d.
\label{eq:conditions}
\EEQ
Then the following holds:
\BEA
\lim_{N\to\infty} \mathbb{E} | {\widehat f\left( \vx \right) - f\left( \vx \right)} | = 0 \\
\lim_{N\to\infty}  \mathbb{E} | {\widehat f\left( \vx \right) - f\left( \vx \right)} |^2 = 0
\EEA
\end{theorem}
The above theorem states that LGDE is asymptotically unbiased and L2-consistent.

\section{LGDE-based Estimators for Entropy and Mutual Information} \label{sec:lge}

We now introduce our estimators for entropy and mutual information that are inspired by the local density estimation approach defined in the previous section.

Let  us again consider $N$ i.i.d samples $(\mathcal{X},\mathcal{Y}) = \{(\vx,\vy)^{(i)}\}_{i=1}^N$ drawn from an unknown joint distribution $f_{XY}$, where $\vx$ and $\vy$ are random vectors of dimensionality $d$ and $b$, respectively. 
Let us construct the following estimators for entropy,
\BEA \label{eq:ent_esti}
\widehat H\left( {\mathbf{x}} \right) &=& -\frac{1}{N}\sum\limits_{i = 1}^N {\log \widehat f\left( {{{\mathbf{x}}_i}} \right)}, 
\EEA
and mutual information
\BEQ\label{eq:mi_esti}
\widehat I\left( {{\mathbf{x}}:{\mathbf{y}}} \right) = \frac{1}{N}\sum\limits_{i = 1}^N {\log \frac{{\widehat f\left( {{{\mathbf{x}}_i},{{\mathbf{y}}_i}} \right)}}{{\widehat f\left( {{{\mathbf{x}}_i}} \right)\widehat f\left( {{\vy_i}} \right)}}} 
\EEQ
where $\widehat f(\vx)$, $\widehat f(\vy)$, $\widehat f(\vx, \vy)$ are the local Gaussian density estimators for $f_X(\vx)$, $f_Y(\vy)$, $f_{XY}(\vx,\vy)$ respectively, defined in the previous section. 

Recall that the entropy and mutual information can be written as appropriately defined expectations; see Eqs.~\ref{eq:naiveE} and ~\ref{eq:naiveI}. Then the proposed estimator simply replaces the expectation by the sample averages, and then plugs in density estimators from Section~\ref{sec:lld} into those expectations.

The next two theorems state that the proposed estimators are asymptotically unbiased. 

\begin{theorem}[\small{Asymptotic Unbiasedness of Entropy Estimator}]\label{theo:ent_consist} If the conditions in Eq.~\ref{eq:conditions} hold, then the entropy estimator given by Eq.~\ref{eq:ent_esti} is asymptotically unbiased, i.e., 
\BEQ
\lim_{N\to \infty} \mathbb{E} {\widehat H\left( {\mathbf{x}} \right)}  = H(\vx)
\label{eq:th1}
\EEQ 
\end{theorem}

\begin{theorem}[\small{Asymptotic Unbiasedness of MI Estimator}]\label{theo:mi_consist}~~~
If the conditions in Eq.~\ref{eq:conditions} hold, then the mutual information estimator given by Eq.~\ref{eq:mi_esti} is asymptotically unbiased:
\BEQ
\lim_{N\to \infty} \mathbb{E} {\widehat I\left( {\mathbf{x}} : \vy \right)}  = I(\vx:\vy)
\label{eq:th2}
\EEQ
\end{theorem}
We provide the proofs of the above theorems in the next section.

\section{Proofs of the Theorems} \label{sec:proof}
Before getting to the actual proofs, we first introduce the Lebesgue's dominated convergence theorem.

\begin{theorem}[Lebesgue dominated convergence theorem]\label{theo:lebesgue}
Let $\{f_N\}$ be a sequence of functions, and assume this sequence  converges point-wise to a function $f$, i.e., $f_N(\vx) \to f(\vx)$ for any $\vx\in R^d$. Furthermore, let us assume that $f_N$ is dominated by an integrable function $g$, e.g., we have for any $\vx$
\BEA
|f_N(\vx)| \le g(\vx) \nonumber 
\EEA
Then we have
\BEA
{\lim _{N \to \infty }}\int_{{\mathbf{x}} \in {\mathbf{X}}} {\left| {{f_N}\left( {\mathbf{x}} \right) - f\left( {\mathbf{x}} \right)} \right|d{\mathbf{x}}}  = 0 \nonumber
\EEA
\end{theorem}

\subsection{Proof of Theorem~\ref{theo:ent_consist}}
Consider $N$ i.i.d.\ samples $\left\{ {{{\mathbf{x}}^{\left( i \right)}}} \right\}_{i = 1}^N$ drawn from the probability density $f(\vx)$, and let $F_N(\vx)$ denote the empirical cumulative distribution function. 

Let us define the following two quantities:
\BEA
{H_1} &=&  - \frac{1}{N}\sum\limits_{i = 1}^N {\ln \mathbb{E}\widehat f( {{{\mathbf{x}}_i}} )}\\
{H_2} &=&  - \frac{1}{N}\sum\limits_{i = 1}^N {\ln f( {{{\mathbf{x}}_i}} )} 
\EEA
Then we have,
\BEA \label{eq:deco}
&~&\mathbb E | {\widehat H ( {\mathbf{x}} ) - H( {\mathbf{x}} )} |  \nonumber \\
 && = \mathbb E | {( {\widehat H - {H_1}} ) + ( {{H_1} - {H_2}} ) + ( {{H_2} - H} )} |  \nonumber \\
  && \le \mathbb E | {\widehat H - {H_1}} | + \mathbb E | {{H_1} - {H_2}} | + \mathbb E | {{H_2} - H} | 
\label{eq:deco}
\EEA
We now procced to show that each of the terms in Eq.~\ref{eq:deco} individually converges to  $0$ in the limit $N \to \infty$, which will then yield Eq.~\ref{eq:th1}.
First, we note that according to the mean value theorem, for any $\vx$, there exist $t_{\vx}$ and $t'_{\vx}$ in $(0,1)$, such that
\BEA \label{eq:taylor1}
&&\ln \widehat f\left( {\mathbf{x}} \right) = \ln \mathbb E\widehat f\left( {\mathbf{x}} \right) +  \\
&&\left( {\widehat f\left( {\mathbf{x}} \right) - \mathbb E\widehat f\left( {\mathbf{x}} \right)} \right){\ln' }\left( {t_{\vx}\widehat f\left( {\mathbf{x}} \right) + \left( {1 - t_{\vx}} \right)\mathbb E\widehat f\left( {\mathbf{x}} \right)} \right) \nonumber
\EEA
and 
\BEA \label{eq:taylor2}
&~&\ln \mathbb E\widehat f\left( {\mathbf{x}} \right) =  \ln f\left( {\mathbf{x}} \right) + \\
&&\left( {\mathbb E\widehat f\left( {\mathbf{x}} \right) - f\left( {\mathbf{x}} \right)} \right){\ln'}\left( {t'_{\vx}f\left( {\mathbf{x}} \right) + \left( {1 - t'_{\vx}} \right)\mathbb E\widehat f\left( {\mathbf{x}} \right)} \right) \nonumber
\EEA

For the first term in Eq.~\ref{eq:deco}, we use Eq.~\ref{eq:taylor1} to obtain
\begin{eqnarray}\label{eq:h1}
&&  \mathbb E\left| {\widehat H - {H_1}} \right| ~~~~~~~~\nonumber \\ 
   &&= \mathbb E\left| {\int {[ {\ln \widehat f\left( {\mathbf{x}} \right) - \ln \mathbb E\widehat f\left( {\mathbf{x}} \right)} ]d{F_N}\left( {\mathbf{x}} \right)} } \right| \hfill \nonumber \\
   &&= \mathbb E\left| {\int {\frac{{| {\widehat f( {\mathbf{x}} ) - \mathbb E\widehat f( {\mathbf{x}} )} |}}{{t_{\vx}\widehat f\left( {\mathbf{x}} \right) + \left( {1 - t_{\vx}} \right)\mathbb E\widehat f\left( {\mathbf{x}} \right)}}d{F_N}\left( {\mathbf{x}} \right)} } \right| \hfill  \nonumber \\
   &&\le \frac{1}{{1 - t}}\mathbb E\left| {\int {\frac{{| {\widehat f\left( {\mathbf{x}} \right) - \mathbb E\widehat f\left( {\mathbf{x}} \right)} |}}{{\mathbb E\widehat f\left( {\mathbf{x}} \right)}}d{F_N}\left( {\mathbf{x}} \right)} } \right| \hfill \nonumber \\
   &&= \frac{1}{{1 - t}}\mathbb E\left( {\frac{1}{N}\sum\limits_{i = 1}^N {\frac{{| {\widehat f\left( {{{\mathbf{x}}_i}} \right) - \mathbb E\widehat f\left( {{{\mathbf{x}}_i}} \right)} |}}{{\mathbb E\widehat f\left( {{{\mathbf{x}}_i}} \right)}}} } \right) \hfill  \nonumber\\
   &&= \frac{1}{{1 - t}}\mathbb E\left( {\mathbb E\left( {\frac{{| {\widehat f\left( {\mathbf{u}} \right) - \mathbb E\widehat f\left( {\mathbf{u}} \right)} |}}{{\mathbb E\widehat f\left( {\mathbf{u}} \right)}}} \right)|{\mathbf{x}} = {\mathbf{u}}} \right) \hfill \nonumber\\
   &&= \frac{1}{{1 - t}}\int {| {\widehat f\left( {\mathbf{u}} \right) - \mathbb E\widehat f\left( {\mathbf{u}} \right)} |\frac{{\widehat f\left( {\mathbf{u}} \right)}}{{\mathbb E\widehat f\left( {\mathbf{u}} \right)}}} d{\mathbf{u}} \hfill 
\end{eqnarray}
where $t$ is the maximum value among all $t_{\vx}$. Using Theorem~\ref{theo:lgd_consist}, we have ${| {\widehat f\left( {\mathbf{u}} \right) - \mathbb E\widehat f\left( {\mathbf{u}} \right)}  |} \to 0$ as $N\to\infty$. Furthermore, it is possible to show that $\exists N_0$, so that for any $N>N_0$ one has $ { | {\widehat f\left( {\mathbf{u}} \right) - \mathbb E\widehat f\left( {\mathbf{u}} \right)} |\frac{{\widehat f\left( {\mathbf{u}} \right)}}{{\mathbb E\widehat f\left( {\mathbf{u}} \right)}}} <2 f\left( {\mathbf{u}} \right)$.  
Thus, using Theorem~\ref{theo:lebesgue}, we obtain 
\BEQ \label{eq:h1_2}
\lim_{N \to \infty}\mathbb E| {H - {H_1}} | = 0
\EEQ
Similarly, using Eq.~\ref{eq:taylor2}, $\mathbb E\left| {{H_1} - {H_2}} \right|$ can be written as 
\begin{eqnarray}\label{eq:h2}
&&\mathbb E\left| {{H_1} - {H_2}}  \right| ~~~~~~~~\nonumber \\
&&= \mathbb E\left| {\int {[ {\ln \mathbb E\widehat f\left( {\mathbf{x}} \right) - \ln f\left( {\mathbf{x}} \right)} ]d{F_N}\left( {\mathbf{x}} \right)} } \right| \nonumber \\
&&=\mathbb E\left| {\int {\frac{{| {\mathbb E\widehat f\left( {\mathbf{x}} \right) - f\left( {\mathbf{x}} \right)} |}}{{t'_{\vx}f\left( {\mathbf{x}} \right) + \left( {1 - t'_{\vx}} \right)\mathbb E\widehat f\left( {\mathbf{x}} \right)}}d{F_N}\left( {\mathbf{x}} \right)} } \right| \hfill \nonumber \\
&&\le \frac{1}{t'}\mathbb E\left| {\int {\frac{{| {\mathbb E\widehat f\left( {\mathbf{x}} \right) - f\left( {\mathbf{x}} \right)} |}}{{f\left( {\mathbf{x}} \right)}}d{F_N}\left( {\mathbf{x}} \right)} } \right| \nonumber \\
&&= \frac{1}{t'}\mathbb E\left( {\frac{1}{N}\sum\limits_{i = 1}^N {\frac{{| {\mathbb E\widehat f\left( {{{\mathbf{x}}_i}} \right) - f\left( {{{\mathbf{x}}_i}} \right)} |}}{{f\left( {{{\mathbf{x}}_i}} \right)}}} } \right) \nonumber \\
&&= \frac{1}{t'}\int {f\left( {\mathbf{x}} \right)\frac{{| {\mathbb E\widehat f\left( {\mathbf{x}} \right) - f\left( {\mathbf{x}} \right)} |}}{{f\left( {\mathbf{x}} \right)}}d{\mathbf{x}}}  \nonumber \\
  &&= \frac{1}{t'}\int {| {\mathbb E\widehat f\left( {\mathbf{x}} \right) - f\left( {\mathbf{x}} \right)} |d{\mathbf{x}}}  
\end{eqnarray}
where $t'$ is the minimum value among all $t'_{\vx}$.

Invoking Theorem~\ref{theo:lgd_consist} again, we observe that the last term in Eq.~\ref{eq:h2} ${| {\mathbb E\widehat f\left( {\mathbf{x}} \right) - f\left( {\mathbf{x}} \right)} | \to 0}$ as $N \to \infty$, and is bounded by $2f(\vx)$ for sufficiently large $N$ (e.g., when when $\widehat f(\vu)$ and $\mathbb E \widehat f(\vu)$ are sufficiently close). Therefore, by Theorem~\ref{theo:lebesgue}, we have
\BEA \label{eq:h2_1}
\lim_{N \to \infty} \mathbb E\left| {{H_1} - {H_2}} \right|  = 0 
\EEA
Finally, for the last term in Eq.~\ref{eq:deco}, we note that
\BEA \label{eq:h2_2}
  \mathbb E {{H_2}}  =  - \frac{1}{N}\mathbb E\sum\limits_{i = 1}^N {\ln f\left( {{{\mathbf{x}}_i}} \right)} =   \mathbb E [-\ln f\left( {\mathbf{x}} \right)]
\EEA 
Thus,  $ \mathbb E {{H_2}}$ is simply the entropy in Definition~\ref{def:entropy_def}; see Eq.~\ref{eq:naiveE}. Therefore, 
\BEA \label{eq:h3}
\lim_{N \to \infty} \mathbb E\left| {{H_2} - H} \right|=0
\EEA
Combining Eqs.~\ref{eq:h1_2},~\ref{eq:h2_1},~\ref{eq:h3} and~\ref{eq:deco}, we arrive at Eq.~\ref{eq:th1}, which concludes the proof. 

\subsection{Proof of Theorem~\ref{theo:mi_consist}}
For mutual information estimation, we use Eq.~\ref{eq:ent_mi} to get

\BEA
\label{eq:mi_deco}
  \mathbb E | {\widehat I\left( {{\mathbf{x}}:{\mathbf{y}}} \right) - I\left( {{\mathbf{x}}:{\mathbf{y}}} \right)} | &\le& \mathbb E | {H\left( {\mathbf{x}} \right) - \widehat H\left( {\mathbf{x}} \right)} |~~~~~~~~\nonumber\\
&+& \mathbb E | {H\left( {\mathbf{y}} \right) - \widehat H\left( {\mathbf{y}} \right)} | ~~~~~~~~ \nonumber\\
&+& \mathbb E | {H\left( {{\mathbf{x}},{\mathbf{y}}} \right) - \widehat H\left( {{\mathbf{x}},{\mathbf{y}}} \right)} | ~~~~~~~~
\EEA
Using Theorem~\ref{theo:ent_consist}, we see that all three terms on the right hand side in Eq.~\ref{eq:mi_deco} converge to zero as $N \to \infty$, therefore $\lim_{N \to \infty}  \mathbb E | {\widehat I\left( {{\mathbf{x}}:{\mathbf{y}}} \right) - I\left( {{\mathbf{x}}:{\mathbf{y}}} \right)} | = 0$, thus concluding the proof.

\section{Experiments} \label{sec:exp}
\subsection{Implementation Details}
Our main computational task is to maximize the local likelihood function in Eq.~\ref{eq:llf}. Since computing the second term on the right hand side of Eq.~\ref{eq:llf} requires integration that can be time-consuming, we choose the kernel function $K(\cdot)$ to be a Gaussian kernel, $\mathbf{K}_{\mathbf{H}} ( \mathbf{t} - \mathbf{x}) = {{\mathcal{N}_d}( {{\mathbf{t}};{\mathbf{x}},{\mathbf{H}}})}$ so that the integral can be performed analytically, yielding

\BEA
\int {{{\mathbf{K}}_{\mathbf{H}}}\left( {{\mathbf{t}} - {\mathbf{x}}} \right){\mathcal N_d}\left( {{\mathbf{t}};\bm{\mu} ,\bm{\Sigma} } \right)} d{\mathbf{t}} = {\mathcal{N}_d}\left( {{\mathbf{x}};\bm \mu ,{\mathbf{H}} +\bm \Sigma } \right) 
\EEA
Thus, Eq.~\ref{eq:llf} reduces to 
\BEA
\label{eq:llf2}
\mathcal L\left( {{\mathbf{x}},\bm \mu ,\bm \Sigma } \right)  &=& \frac{1}{N}\sum\limits_{i = 1}^N {{\mathcal{N}_d}\left( {{{\mathbf{x}}_i};{\mathbf{x}},{\mathbf{H}}} \right)\log } {\mathcal{N}_d}\left( {{{\mathbf{x}}_i};\bm \mu ,\bm \Sigma } \right) \nonumber \\
&-& {\mathcal{N}_d}\left( {{\mathbf{x}};\bm \mu ,{\mathbf{H}} +\bm \Sigma } \right) 
\EEA
Maximizing Eq.~\ref{eq:llf2} is a constrained non-convex optimization problem with the condition that the covariance matrix $\bm \Sigma$ is positive semi-definite. We use Cholesky parameterization to enforce the positive semi-definiteness of $\bm \Sigma$, which allows to reduce our constrained optimization problem into an unconstrained one. Also, since we would like to preserve the local structure of the data, we select the bandwidth to be close to the distance between pair of k-nearest points (averaged over all the points).

We use Newton-Ralphson method to do the maximization although the function itself is not exactly concave. The full algorithm for our estimator is given in Algorithm~\ref{alg:mie} which takes Algorithm~\ref{alg:ee} as a subroutine. Note that in Algorithm~\ref{alg:ee}, the Wolfe condition is a set of inequalities in performing quasi-Newton methods~\citep{wolfe1969convergence}.
\begin{algorithm}[ht]
\caption{\textbf{Mutual Information Estimation with Local Gaussian Approximation}}
\label{alg:mie}
\begin{algorithmic}
\State{\textbf{Input:} points $(\vx,\vy)^{(1)},(\vx,\vy)^{(2)},...,(\vx,\vy)^{(N)}$}
\State{\textbf{Output:} $\widehat I(\vx;\vy)$}
\State{Calculate entropy $\widehat H(\vx)$ using samples $\vx^{(1)}$, $\vx^{(2)}$...,$\vx^{(N)}$}
\State{Calculate entropy $\widehat H(\vy)$ using samples $\vy^{(1)}$, $\vy^{(2)}$...,$\vy^{(N)}$}
\State{Calculate joint entropy $\widehat H(\vx,\vy)$ using  input samples $(\vx,\vy)^{(1)},(\vx,\vy)^{(2)},...,(\vx,\vy)^{(N)}$}
\State{Return estimated mutual information $\hat I = \widehat H(\vx) + \widehat H(\vy) - \widehat H(\vx,\vy)$}
\end{algorithmic}

\end{algorithm}

\begin{algorithm}[htbp]
\caption{\textbf{Entropy Estimation with Local Gaussian Approximation}}
\label{alg:ee}
\begin{algorithmic}

\State{\textbf{Input:} points $\vu^{(1)},\vu^{(2)},...,\vu^{(N)}$}
\State{\textbf{Output:} $\widehat H(\vu)$}
\State{Initialize $\widehat H(\vu) = 0$}
\For{each point $\vx^\ii$}
\State{initialize $\bm \mu = \bm \mu_0$, $\bm L = \bm L_0$}
\While{not $\mathcal L(\vx^\ii, \bm \mu, \bm \Sigma = \bm L * \bm L^T)$ converge}
\State{Calculate $\mathcal L(\vx^\ii, \bm \mu, \bm \Sigma = \bm L * \bm L^T)$}
\State{Calculate gradient vector $\mathbf G$ of $\mathcal L(\vx^\ii, \bm \mu, \bm \Sigma =$} 
\State{$\bm L * \bm L^T)$, with respect to $\bm \mu$, $\bm L$}
\State{Calculate Hessian matrix of $\mathbf{H}$ of $\mathcal L(\vx^\ii, \bm \mu, \bm \Sigma =$} 
\State{$\bm L * \bm L^T)$, with respect to $\bm \mu$, $\bm L$}
\State{Do Hessian modification to ensure the positive}
\State{semi-definiteness of $\mathbf{H}$}
\State{Calculate descent direction $\mathbf D = -\alpha \mathbf H^{-1} \mathbf G$,} 
\State{where we compute $\alpha$ to satisfy Wolfe condition} 
\State{Update $\bm \mu, \bm L$ with $(\bm \mu, \bm L) + \mathbf D$}
\EndWhile
\State{$\widehat f(\vx^\ii) = \mathcal N(\vx;\bm \mu, \bm \Sigma = \bm L * \bm L^T)$}
\State{$\widehat H(\vu) = \widehat H(\vu) - \frac{\log f(\vx^\ii)}{N} $}
\EndFor
\end{algorithmic}
\end{algorithm}

In a single step, evaluating the gradient and Hessian in Algorithm~\ref{alg:ee} would take $O(N)$ time because Eq.~\ref{eq:llf} is a summation over all the points. However, for points that are far from the current point $\vx^\ii$, the kernel weight function is very close to zero and we can ignore those point and do the summation only over a local neighborhood of $\vx^\ii$.

\subsection{Experiments with synthetic data}
\paragraph{Functional relationships} We test our MI estimator for near-functional relationships of form $Y = f(X) + \mathcal{U}(0,\theta)$, where  $\mathcal{U}(0,\theta)$ is the uniform distribution over the interval $(0,\theta)$, and $X$ is drawn randomly uniformly from $[0,1]$. Similar relationships were studied in~\citep{reshef}, \citep{kinney} and \citep{shuyang2015AISTATS}.

\begin{figure}[ht] 
   \centering
 \includegraphics[width=0.5\textwidth]{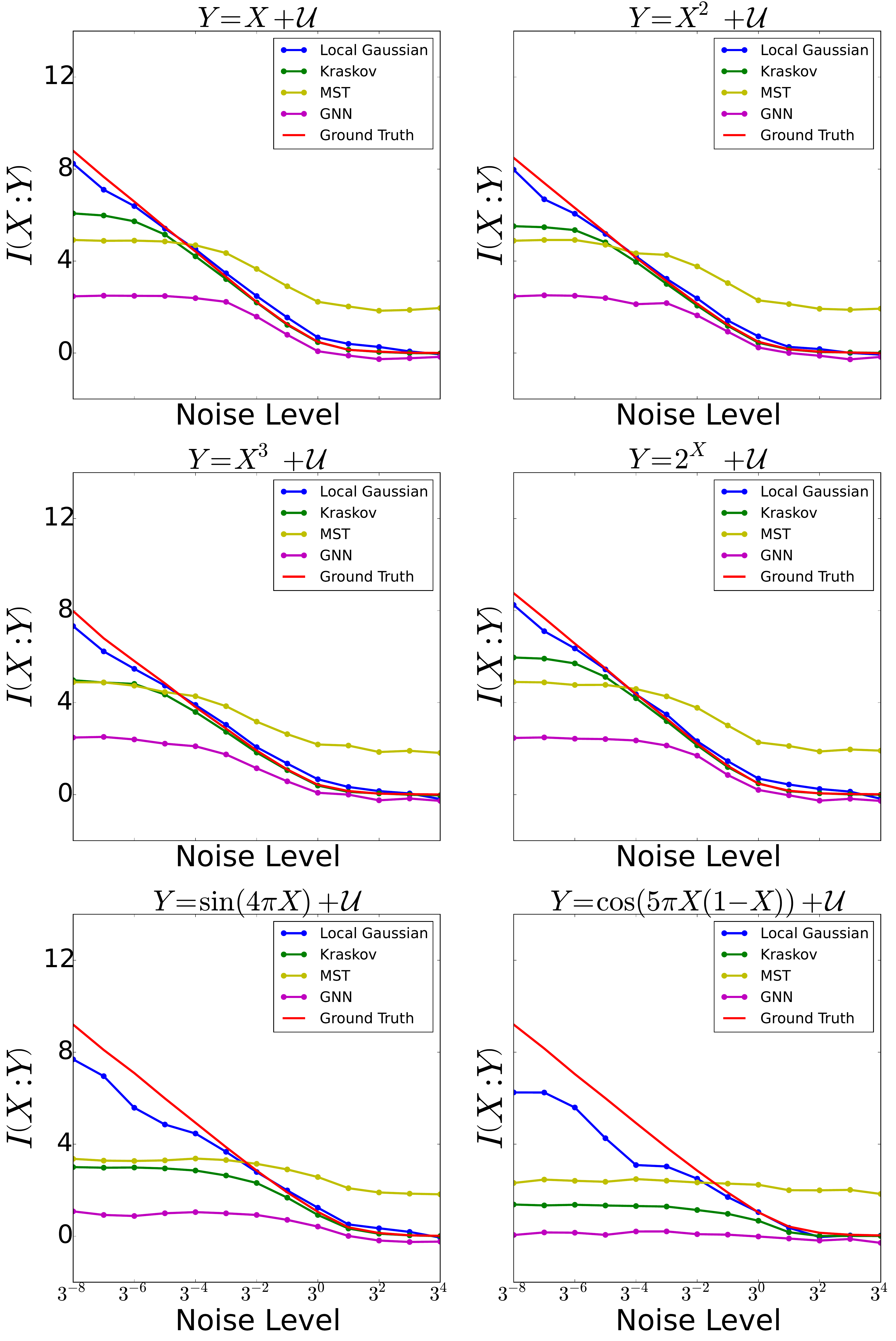} 

   \caption{Functional relationship test for mutual information estimators. The horizontal axis is the value of $\theta$ which controls the noise level; the vertical axis is the mutual information in nats. For the Kraskov and GNN estimators we used nearest neighbor parameter $k=5$. For the local Gaussian estimator, we choose the bandwidth to be the distance between a point and its $5$th nearest neighbor.} 
   \label{fig:noise_relationship} 
\end{figure}

We compare our estimator to several baselines that include the kNN estimator proposed by~\citep{kraskov}, an estimator based on generalized nearest-neighbor graphs (GNN)~\citep{GNNG}, and minimum spanning tree method (MST)~\citep{MST_MI}. We evaluate those estimators for six different functional relationships as indicated in Figure~\ref{fig:noise_relationship}. We use $N=2500$ sample points for each relationship. To speed up the optimization, we limited the summation in Eq.~\ref{eq:llf2} to only $k$ nearest neighbors, thus reducing the computational complexity from $O(N)$ to $O(k)$ in every iteration step of Algorithm~\ref{alg:ee}. 

One can see from Fig.~\ref{fig:noise_relationship} that when $\theta$ is relatively large, all methods except MST produce accurate estimates of MI. However, as one decreases $\theta$, all three baseline estimators start to significantly underestimate  mutual information. In this low-noise regime, our proposed estimator outperforms the baselines, at times by a significant margin. Note also that all the estimators, including ours, perform relatively poorly for highly non-linear relationships (the last row in Figure~\ref{fig:noise_relationship}). According to our intuition, this happens when the scale of the non-linearity becomes sufficiently small, so that the linear approximation of the relationship around the local neighborhood of each sample point does not hold. Under this scenario, accuracy can be recovered by adding more samples.

\section{Related Work}
{\bf Mutual Information Estimators} Recently, there has been a significant amount of work on estimating information-theoretic quantities such as entropy, mutual information, and divergences, from i.i.d. samples. Methods include k-nearest-neighbors~\citep{kNN_naive}, ~\citep{kraskov}, ~\citep{GNNG}, ~\citep{poczos2012nonparametric}; minimum spanning trees~\citep{MST_MI}; kernel density estimate~\citep{moon1995estimation},~\citep{singh_generalized_2014}; maximum likelihood density ratio~\citep{densityratio}; ensemble methods~\citep{moon_ensemble_divergence}, ~\cite{Sricharan_ensemble}, etc. As pointed our earlier, all of those methods underestimate the mutual information when  two variables have strong dependency. \citep{shuyang2015AISTATS} addressed this shortcoming by introducing a local non-uniformity correction, but their estimator depended on a heuristically defined threshold parameter and lacked performance guarantees.

{\bf Density Estimation and Boundary Bias} Density estimation is a classic problem in statistics and machine learning. Kernel density estimation and k-nearest-neighbor density estimates are the two most popular and successful non-parametric methods. However, it has been recognized that these non-parametric techniques often suffer from the problem of so-called ``boundary bias''. Researchers have proposed a variety of methods to overcome the bias, such as the reflection method~\citep{schuster1985incorporating},~\citep{silverman1986density}; the boundary kernel method~\citep{zhang2000nonparametric}, the transformation method~\citep{marron1994transformations}, the pseudo-data method~\citep{cowling1996pseudodata} and others. All these methods are useful in some particular settings. But when it comes to mutual information estimation, how can we choose the most efficient one to use? It seems that local likelihood method~\citep{hjort1996locally}, ~\citep{loader1996local}, is a good choice for estimating the mutual information due to its ability to detect the boundary without any prior knowledge. Previous studies have already proven the power of \textit{local regression}, which can automatically overcome the boundary bias. Methods based on \textit{local likelihood} estimation has traditionally attracted less attention due to their computational complexity. However, advances in computational power allow us to re-consider this class of method.

\section{Conclusion and Future Work }
Past research on mutual information estimation has mostly focused on distinguishing weak dependence from independence. However, in the era of big data, we are often interested in highlighting the strongest dependencies among a large number of variables. When those variables are highly inter-dependent, traditional non-parametric mutual information estimators  fail to accurately estimate the value due to the boundary bias. 

We have addressed this shortcoming by introducing a novel semi-parametric method for estimating entropy and mutual information based on local Gaussian approximation of the unknown density at the sample points. We demonstrated that the proposed estimators are asymptotically unbiased. We also showed empirically that the proposed estimator has a superior performance compared to a number of popular baseline methods, and can accurately measure strength of the relationship even for strongly dependent variables, and limited number of samples. 

There are several potential avenues for future work. First of all, we would like to validate the proposed estimator in higher-dimensional settings. In principle, the approach is general and can be applied in any dimensions. However, the optimization procedure may be computational expensive in higher dimensions, since the number of parameters scales as $O(d^2)$ with dimensionality $d$. An intuitive solution would be to initialize the parameters with the results obtained from the close points, which can facilitate convergence. 

Another interesting issue is the bandwidth selection, which is an  important problem in general density estimation problems. If the bandwidth is too large, the local Gaussian assumption may not be valid, whereas very small bandwidth will result in non-smooth densities. Ideally, we would like to choose the bandwidth in a way that preserves the local Gaussian structure in the neighborhood of each point. Another interesting extension would be choosing the bandwidth adaptively for each point. 

Finally, while here we have focused on the asymptotic unbiasedness of the proposed estimator, it will be very valuable to establish theoretical results about the convergence rates of the estimators, as well as its variance in the large sample limit.

{\em Note added in proof:} We have become aware of a very recent paper on non-parametric entropy estimation ~\citep{another_LGA} that is also based on local Gaussian approximation. Specifically, the kpN estimator suggested in ~\citep{another_LGA} fits a Gaussian distribution with the empirical mean and covariance matrix of the p-nearest neighbors of each point, and then uses this distribution to approximate the probability mass contained in the kNN ball centered at that point. Despite obvious similarities, we note that our approach is based on a local minimization of the Kullback-Leibler distance between the true and the approximating Gaussian densities, whereas the kpN estimator works by fitting a truncated Gaussian distribution. As a result, we are able to derive formal performance guarantees, thus making our approach theoretically better grounded.

\subsubsection*{Acknowledgements}
This research was supported in part by DARPA grant No. W911NF--12--1--0034.

\bibliographystyle{plainnat}
{ \small
\bibliography{sygao}{}
}

\end{document}